\documentclass[12pt]{article}
\usepackage[authordate, backend=biber]{biblatex-chicago}
\usepackage[nopatch]{microtype}
\usepackage[margin=1in]{geometry}
\usepackage{booktabs}
\usepackage{algorithm}
\usepackage{algpseudocode}
\usepackage{amsmath}
\usepackage{amssymb}
\usepackage{amsfonts}
\usepackage{dsfont}
\usepackage[table]{xcolor}
\usepackage{graphicx}
\usepackage{tikz}
\usepackage{subcaption}
\usepackage{amsthm}
\usepackage{hyperref}
\theoremstyle{definition}

\usepackage{bm}

\newtheorem{proposition}{Proposition}

\def\Var{{\textrm{Var}}\,}

\def\Pr{{\textrm{Pr}}\,}
\def\E{{\textrm E}\,}

\def\1{{\mathds{1}}}
\def\arg{{\textrm {arg} }\,}
\def\Cov{{\textrm{Cov} }\,}

\def\Supp{{\textrm {Supp} }\,}
\def\Sym{{\textrm {Sym} }\,}

\usepackage{color} 
\definecolor{direct}{HTML}{FF0000}
\definecolor{indirect}{HTML}{FF9999}
\definecolor{dirin}{HTML}{990000}
\definecolor{control}{HTML}{999999}
\definecolor{darkgreen}{RGB}{0,100,0}

\newcommand{\D}{\mathbf{D}}

\newcommand{\RR}{\mathbb{R}}

\newcommand{\X}{\mathbf{X}}
\newcommand{\Y}{\mathbf{Y}}
\newcommand{\Z}{\mathbf{Z}}

\newcommand{\dd}{\mathbf{d}}

\newcommand{\z}{\mathbf{z}}

\newtheorem{condition}{Condition}[section]

\title{On the Foundations of the Design-Based Approach}
\author{
P.M. Aronow\thanks{Statistics \& Data Science and Political Science, Yale University, USA}
\and Austin Jang\thanks{Statistics \& Data Science and Political Science, Yale University, USA}
\and Molly Offer-Westort\thanks{Department of Political Science, University of Chicago. Corresponding author: Molly Offer-Westort; email: mollyow@gmail.com. Authors are listed in alphabetical order.}
}
\date{}

\hypersetup{
  pdftitle={On the Foundations of the Design-Based Approach},
  pdfauthor={P.M. Aronow; Austin Jang; Molly Offer-Westort},
  pdfsubject={arXiv preprint},
  pdfkeywords={design-based inference, causal inference, interference, exposure mappings, SUTVA},
  pdfdisplaydoctitle=true
}

\begin{document}
\maketitle

\begin{abstract}

The design-based paradigm may be adopted in causal inference and survey sampling when we assume Rubin's stable unit treatment value assumption (SUTVA) or impose similar frameworks.
While often taken for granted, such assumptions entail strong claims about the data generating process. 
We develop an alternative design-based approach: 
we first invoke a generalized, non-parametric model that allows for unrestricted forms of interference, such as spillover. 
We define an associated set of inferential targets and discuss their interpretation under SUTVA and a weaker assumption that we call the No Unmodeled Revealable Variation Assumption (NURVA). 
We then reconstruct the standard paradigm, reconsidering SUTVA at the end rather than assuming it at the beginning. 
Despite its similarity to SUTVA, we demonstrate the practical {limitations} of NURVA {alone} for identifying substantively interesting quantities. 
In so doing, we provide clarity on the nature and importance of SUTVA for applied research.
\end{abstract}

\section{Introduction}
In classical survey sampling, design-based and model-based inference differ by the assumed source of stochasticity in the data-generating process.%
\footnote{Here, we are interested in the classical, statistical meaning of design-based inference, although the term often takes on a different meaning in the social sciences. 
For a reference on the social science interpretation of design-based inference, see: \textcite{card2022design}, \textcite{dunning2010design}, and \textcite{kocher2016lines}.}
In model-based inference, randomness comes from largely allegorical notions such as sampling from an infinite super-population or a stochastic error term in the outcome model.
In design-based inference, randomness lies in which units are treated or sampled under the \textit{design} that assigns probabilities to ``various subsets of the finite population'' \parencite{sarndal1978design}.

Design-based inference for causal queries originated with the influential work of \textcite{neyman1923application} and, shortly after, was developed for survey sampling in \textcite{neyman1934two}.
Neyman-type design-based inference provides asymptotically justified estimation under a set of consistency assumptions and mild regularity conditions.\footnote{For a more comprehensive discussion of the history of design-based survey sampling and causal inference, refer to Section 2 of \textcite{aronow2013class}.} %
The framework established by Neyman was extended to include general causal inference problems including from observational data in \textcite{rubin1974}.\footnote{
Fisher-Rosenbaum type estimation is a distinct but related literature to the approach elaborated in this paper that also assumes treatment assignment is the only source of randomness.
For examples in the social sciences, see: \textcite{sinclair2012detecting} and \textcite{kalla2016campaign}.}
One of Rubin's contributions to the model was to coin the stable unit treatment value assumption (SUTVA).
Today, SUTVA is characterized by two conditions.
The first is \textit{no interference}, as Rubin described,  ``If unit $i$ is exposed to treatment $j$, the observed value of $Y$ will be $Y_{ij}$; that is, there is no interference between units (Cox 1958, p. 19) leading to different outcomes depending on the treatments other units received.''
The second is \textit{no hidden treatment variations}; Rubin continues, ``there are no versions of treatments leading to `technical errors' (Neyman 1935).''
These assumptions let us leverage observations across treatment conditions to make causal claims about alternative interventions. 
In experimental and quasi-experimental settings, SUTVA implies that there are $N{\times}K$ potential outcomes, where $N$ is the number of observations and $K$ is the number of treatment levels, and each observation can take on $K$ potential outcomes.
In the survey sampling context, we invoke similar assumptions. 
Suppose that we survey $n$ of $N$ units in a finite population but we wish to make claims about what would have been observed had we surveyed the entire population.
To do so, we often implicitly rely on a SUTVA-like assumption that there exist only $N$ observable potential outcomes, one associated with each unit of the population.\footnote{See the fixed outcome model in classical survey sampling associated with \textcite{sarndal1978design} or \textcite{kish1965survey}.}
This implies that for every observation we assume that who else, and how, we survey does not affect observable potential outcomes.
While often taken for granted, this is a strong assumption{, and will have implications for the generalizability of causal quantities as well}.

{
There have been a number of advances in recent years for the design-based approach to statistical inference, including for instrumental variables estimation \parencite{ borusyak2024design}, regression adjustment \parencite{middleton2018unified}, and interference both spatially \parencite{leung2020treatment} and temporally \parencite{wang2021causal}.
Additional recent design-based work develops estimation under relaxations of SUTVA in specific settings 
\parencite{zhao2022reconciling, gao2023causal, lu2025design, heng2025design}. 
More general methods for randomization tests under interference \parencite{basse2019randomization} and Riesz estimators \parencite{harshaw2022design} have also been proposed.}

We {contribute to this literature by proposing} a new design-based framework for analyzing randomized trials and survey sampling. 
Departing from the usual reliance on Rubin's SUTVA, our approach begins with a generalized, non-parametric model that permits arbitrary forms of interference. 
{Because unit-level potential outcomes are not generally well-defined under interference or hidden versions of treatment, we introduce design-conditional \textit{expected} potential outcomes and their averages and design-conditional contrasts as an alternative (Equations~\ref{eqn:epo}, \ref{eqn:aepo}, \ref{eqn:eed}, and \ref{eqn:aeed}). 
Our defined contrasts coincide with the usual average treatment effects when SUTVA holds, but these targets remain well-defined as combinations of potential outcomes, even under interference or misspecification of exposures---as long as the \textit{design} is known. 
We introduce a weaker, design-conditional variant of SUTVA that we call the no unmodeled revealable variation assumption (NURVA) (Condition~\ref{condition:nurva}) that requires stability only over the support of the design. 
When NURVA holds but SUTVA does not, contrasts remain causal under the design; but only SUTVA permits off‑support interpretations. }

{ 
This framework enables researchers to conduct more faithful analyses of complex data-generating processes that reflect real-world interactions. 
While off-support generalizations may interest policymakers and applied researchers, our reframing separates design‑identified quantities from those targets. %
}

In Section \ref{sec:examples}, we provide {two} stylized examples illustrating our framework. 
In Section \ref{section:setting}, we establish the key elements of the analysis---design, population, and exposure mapping. 
In Section \ref{section:inferential}, we define a set of new inferential targets for the design-based setting. 
In Section \ref{sec:interpretation}, we discuss the interpretation of these targets under Rubin's SUTVA as well as under NURVA. 
In Section \ref{sec:generalizing}, we discuss implications {of} SUTVA and NURVA for external validity.
In Section \ref{section:estimation}, we discuss implications for estimation. 
In Section \ref{section:conclusion}, we conclude.

\section{Stylized Examples}
\label{sec:examples}

To motivate our discussion, we use {two} stylized versions of common research designs (a survey and an experiment {in a setting where there is potential peer influence}) to demonstrate the flexibility and generality of the proposed framework.
Additional toy models that demonstrate important, but more particular, pathologies will be discussed later.

\paragraph{Example 1: Rebel group survey. } 
A researcher has a list of 10 historical rebel groups in Colombia and asks: what is the average number of reported members today? 
To answer this question{,} they randomly order the 10 groups, survey the first three, and record reported former members; unsurveyed groups are coded as having no members.
Unsurveyed rebel groups can be coded arbitrarily (e.g., 0 or -99); this is inconsequential for defining the target or for estimation.


\paragraph{Example {2: School} experiment.} 
A researcher is interested in the effect of a pro-volunteering program on volunteering among a population of high school students. 
In the experimental design, they assign one-fifth of the students to the pro-volunteering arm under uniform random assignment. 
For each student, the researcher then measures whether they volunteered.

\section{Set-Up}
\label{section:setting}

{Following the frameworks of \textcite{neyman1923application} and \textcite{sarndal1978design}, we formalize the design in terms of a probability triple $(\Omega, S, P)$, where the sample space, $\Omega$, consists of all conceivable experimental (or sampling) interventions and each realized intervention, $\omega$, refers to the vector of treatment (or sampling) assignments to all $N$ observations.
The event space is the power set of the sample space, $S = \mathcal{P}(\Omega)$ and we assume that our probability measure satisfies $\Pr[A] = \sum_{\omega \in A} p(\omega), \forall A \in S$, where $p: \Omega \to [0,1]$, with $p(\omega)$ characterizing the probability of the assignment of an intervention.
We then embed potential outcomes per \textcite{rubin1974} and adopt exposure mappings for interference per \textcite{aronow2017estimating} \parencite[see also][]{hudgens2008toward}. }

\paragraph{Design.} 
%
%
%
We define the design space $\mathcal{Z} = \Z(\Omega)$, where \(\Z\) is a bijective random vector $\Z:\Omega \to \RR^N$. %
The \textit{design} is defined as the joint p.m.f. of $\Z$, $f: \RR^N \to [0,1]$. 
{We may define all objects with respect to the design vector $\Z$, removing all notational dependence on $\omega$.} 
We assume that the design is determined by the researcher, grounding our discussion away from concerns about the uncertainty in the assignment procedure. 
More generally, a valid design exists in settings where the researcher is simply aware of the probabilities implied by an assignment process and the sample space, allowing for design-based inference in observational settings. 
{This setup and notation {are}, so far, fairly standard in the design-based setting. 
However, we will use this setup to define novel estimands in Section~\ref{section:inferential}. 
The conceptual differences between the design space and the design will be central to the interpretation of these estimands as discussed in Section~\ref{sec:interpretation}.}

For illustration, we define the design with respect to our stylized examples.
In the rebel group survey, the design is the sampling procedure that the researcher uses to determine the probability of surveying each of the historical rebel groups, which consists of randomly sorting a list of the ten groups and surveying the first three in order. 
One way to formally express this procedure is as follows,
\[f(\z) = \begin{cases}
		\frac{(10-3)!}{10!}
		 & : \z \in \Sym((1,2,3,0,0,0,0,0,0,0)) \\
		0 & : \text{ otherwise }
	\end{cases}\]
where $\Sym(\cdot)$ represents the set/orbit of distinct permutations of a vector. 
The corresponding design space is $\mathcal{Z} \supseteq \{(0,0,0,0,0,0,0,0,0,0)\} \cup \Sym((1,0,0,0,0,0,0,0,0,0)) \cup \Sym((1,2,0,0,0,0,0,0,0,0)) \cup \dots  \cup \Sym( (1,2,3,4,5,6,7,8,9,10))$.%
\footnote{
In our notational shorthand for $\mathcal{Z}$, the number represents the ordinal position in which the indexed group was surveyed with 0 indicating that the group was not surveyed. 
For instance, $\z = (1,2,3,0,0,0,0,0,0,0)$ translates to surveying group one first, group two second, and group three third, with no other group being surveyed. 
This is a different intervention than $\z = (3,2,1,0,0,0,0,0,0,0)$ in which we survey the same three groups, but we surveyed group three first, group two second, and group one third.}

	
In our {school} experiment, the design is the procedure we use to determine the probability that each student receives the treatment arm. 
Letting $\lceil .2 N \rceil $ be the number of observations in treatment, the design is,
	\[f(\z) = 
	\begin{cases}
		\frac{(\lceil .2 N \rceil)! (N- \lceil .2 N \rceil)! }{N!} & : \sum_{k=1}^N z_k = \lceil .2 N \rceil \\
		0 & : \text{otherwise}
	\end{cases}\]
where the corresponding design space is $\mathcal{Z} = \{0,1 \}^N$.

\paragraph{Population.}
\label{sec:population}
 
We follow \textcite{sarndal1978design} in defining the population as the collection of units about which we wish to conduct inference. 
This finite population is indexed by $i = 1,\dots,N$, where indexing is fixed. 
Our estimands are defined over this population on specified outcomes of interest. 
{Unless otherwise specified, generalizations are to estimands defined over this same population.}
We represent the outcomes of interest associated with these units with a real-valued random vector, $\Y$. 
%
The random vector $\Y$ characterizes the \textit{raw potential outcomes}, which can be represented for each individual $i$ as,
    \begin{equation}
        \label{eqn:raw}
        y_i(\z) = Y_i(\Z^{-1}(\z)), \forall i \in 1,\dots ,N, \z \in \mathcal{Z}.
    \end{equation} 
The raw potential outcome $y_i(\z)$ for individual $i$ is what we observe if we perform intervention $\z$ and measure unit $i$'s outcome. 
There are a total of $N \times| \mathcal{Z}|$ raw potential outcomes in the causal setting. 
There are $\sum_{\z \in \mathcal{Z}} \sum_{z_i \in \z} {\1}\{z_i \neq 0\}$ substantively defined raw potential outcomes in the survey sampling setting.\footnote{The term \textit{raw} refers to the fact that we are characterizing the well-defined potential outcome of unit $i$ under the \textit{intervention}, $\z$, as compared to the distribution of individual potential outcomes that may be associated with unit $i$ under a given researcher defined individual level exposure.} 

In our rebel group survey, the population is the 10 historical rebel groups and the outcome is the measured number of reported members.
In our {school} experiment, the population is the $N$ students, and the outcome is an indicator of whether a student volunteered.

\paragraph{Exposure mapping.}
\label{section:exposure}

While the raw potential outcome characterizes the outcome at the most granular level, we are often interested in further characterizing a unit of observation by the ``exposure'' that our intervention has placed them in. 
Examples of exposures in the experimental literature include social influence \parencite{centola2010spread}, spatial exposure \parencite{miguel2004}, or simple direct exposure---what is most commonly considered in experimental settings under SUTVA. 
The \textit{exposure mapping} relates interventions to the exposures (``treatments'') experienced by the units of observation. The exposure mapping was originally defined, named, and applied to the design-based setting in \textcite{aronow2017estimating}.
When exposure mappings are properly specified, i.e., for each unit a given exposure is associated with a single, stable value of the outcome, and if this is satisfied for all units, then exposures align with what \textcite{manski2013identification} denotes as ``effective treatments.'' 

Formally, our exposure mapping for individual $i$ is a function $g_i: \mathcal{Z} \to \RR$. 
Since $g_i$ is a function, each realization of the intervention associates each individual with only one exposure, although this can be easily relaxed to allow for settings where an observation is in multiple exposures, such as in multi-factorial designs. 
Given that we have knowledge of the support of the design vector, $\Supp{(\Z)}$, we can define a function $g_i| _{\Supp (\Z)} : \Supp(\Z) \to \RR $. 
We denote the $N$-length random exposure vector,
 \begin{equation}
    \label{eqn:exposure}
     \D = (g_1(\Z), g_2(\Z),\dots ,g_N(\Z)).
 \end{equation}
We {condition on the support of the design in $g_i| _{\Supp (\Z)}$ because while} there may be some interventions in the design space, $\mathcal{Z}$, in which the exposure is not well-defined, for estimation these are only consequential if they occur with positive probability.

In our rebel group survey, we might code a group as exposed if they had been surveyed and not exposed if they were not. This exposure mapping can be expressed mathematically as,
	\[
	g_i(\z) = \begin{cases}
		1  & : z_i >0 \\
		0 & : \text{ otherwise }
	\end{cases}, \forall i \in \{1,\dots ,10\}, \z \in \Sym((1,2,3,0,0,0,0,0,0,0)).
	\]

In our {school} experiment, we could use an ``individualistic'' exposure mapping, {in which we code an individual as exposed if they were assigned to the program, and not exposed otherwise}, 

    \[
        g_i(\z) = \begin{cases}
	1  & : z_i = 1\\
	0 & : \text{ otherwise }
        \end{cases}, \forall i \in \{1,\dots ,N\},  \z \in \{0,1\}^N.
    \]
Alternatively, {suppose the researcher also has some pre-treatment measure of a network among peers, which may deviate from the true network in its representation of the underlying data generating process.}
Using this pre-treatment measure of the social network, we could define exposures based on whether adjacent nodes were treated \parencite{aronow2017estimating,savje2024causal}. 
One example is:

\begin{equation}
	\label{eqn:network}
	g_i(\z) = \begin{cases}
    	0 \text{ (Control)} & : \text{if } z_i = 0 \text{ and } \theta_i\z = 0, \\
		1 \text{ (Isolated Direct)} & : \text{if } z_i = 1 \text{ and } \theta_i\z = 0, \\
		2 \text{ (Indirect)} & : \text{if } z_i = 0 \text{ and } \theta_i\z > 0, \\
		3 \text{ (Direct \& Indirect)} & : \text{if } z_i = 1 \text{ and } \theta_i\z > 0, \\
	\end{cases}, \forall i \in \{1,\dots ,N\},  \z \in \{0,1\}^N,
\end{equation}
where $\theta_i$ is the $i$th row of the network adjacency matrix.
\begin{figure}[!htbp]
	\centering
	\caption{Network Exposure}
	\label{fig:network_exposure}
	\includegraphics[width=0.45\textwidth]{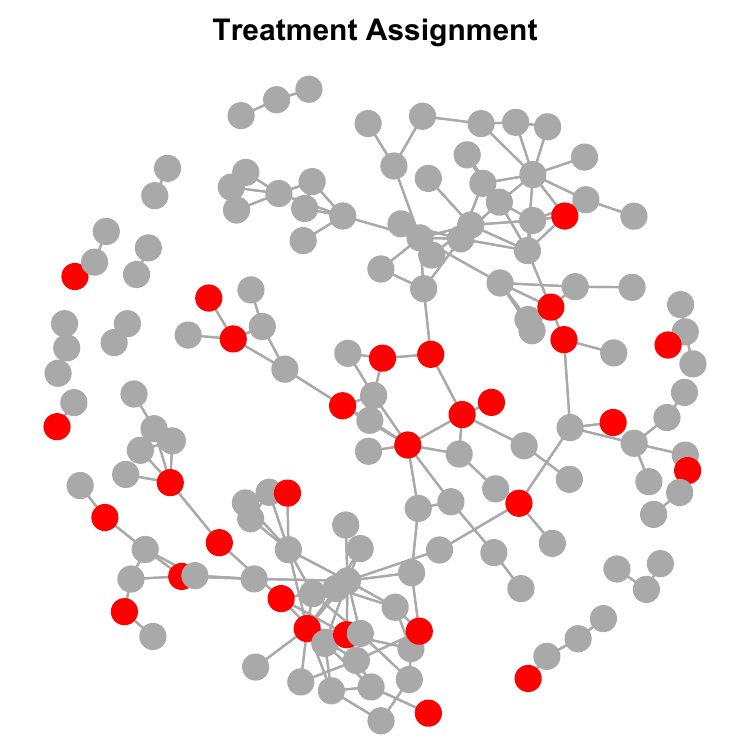}\hspace{0.02\textwidth}$\rightarrow$\hspace{0.02\textwidth}\includegraphics[width=0.45\textwidth]{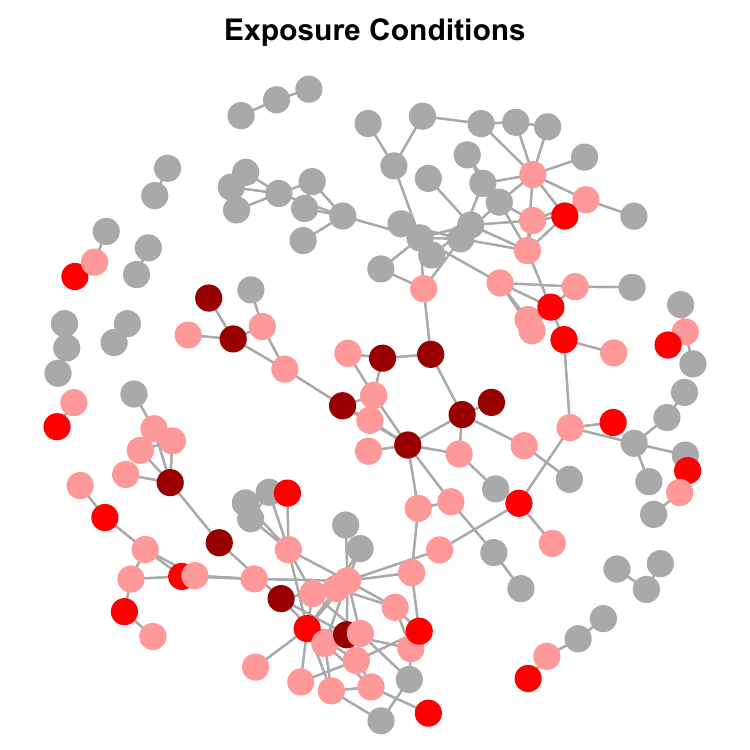}
\end{figure}

Figure \ref{fig:network_exposure} illustrates this mapping: the left panel shows the intervention (treated in red, control in gray); the right panel shows resulting exposures under Equation \ref{eqn:network}: control (gray), isolated direct (red), indirect (pink), and direct plus indirect (maroon).
 
In Section \ref{sec:interpretation}, we present a toy example where the exposure mapping does not match the true underlying decision rule and discuss implications. 
\textcite{savje2024causal} further delineates the use of the exposure mapping to define causal estimands and for making assumptions about our potential outcomes.

\section{Inferential targets}
\label{section:inferential}

{We define new inferential targets for the design-based framework.}
In each experimental or survey sampling setting, we observe only one realization of $(\Z, \Y)$.
For experimental estimands, we first define contrasts between the individual-level outcome {in expectation} under a design when a unit has been exposed to one condition and the outcome {in expectation} when that same unit has been exposed to another. 
We use these individual-level contrasts as building blocks for our eventual inferential targets, averages of these contrasts over units in the population. 
Similarly in the survey sampling setting, we define the individual-level outcome {in expectation} under a design when a unit has been sampled, and then construct estimands as averages across units. 

Note that under this set-up, the ``potential outcome when unit $i$ is in exposure $d$'' is not generally well-defined. 
There may be multiple assignment vectors $\z$ compatible with each exposure $d$, and thus multiple possible raw potential outcomes. 
Most other work assumes this away (most commonly with Rubin's SUTVA, further discussed in Section \ref{sec:interpretation}), which allows for an unambiguous characterization of the mean outcome or average treatment effect. 
Instead, we define alternative targets that coincide with standard targets when SUTVA holds, but remain well-defined when it does not. 
In cases when SUTVA fails, {absent alternative stability assumptions,} these targets are akin to ``pseudo-true'' parameters, values that minimize a population-level objective function under model misspecification \parencite{sawa1978information}. 

\paragraph{Expected Potential Outcome {(EPO).}}
The first target we consider is the \textit{expected potential outcome} (EPO). 
First, we denote the marginal probability that unit $i$ enters exposure $d$ under the design distribution of $\Z$ as,
    \begin{equation}
        \label{eqn:probability}
        \pi_i(d) = \left(\sum_{\z \in \Supp (\Z): g_i(\z) = d} \Pr[\Z = \z] \right)= \Pr[D_i = d].
    \end{equation}

We impose a condition that all observations have positive probability under the design of falling into the relevant exposure condition.  

\begin{condition}{(Individual positivity)}
\label{condition:positivity}
An exposure $d$ satisfies \textit{individual positivity} if there exists a $c$ such that $0 < c \le \pi_i(d) \ \forall i \in \{1, \dots, N \}$. 
\end{condition}

Note that unlike in conventional approaches to inverse probability weighting, in which positivity is a population-level assumption used for estimation, here it is an individual-level assumption also used in defining the estimands. 
More specifically, in conventional inverse probability weighting, an individual may receive a treatment deterministically if, when drawing from the population, a given covariate profile has a non-zero probability of being observed under any treatment condition \parencite{hernan2006estimating, cole2008constructing}. 
In our setting, each individual that we conduct inference over must have a non-zero probability of being in the relevant exposure condition. 
While this may seem like a stringent assumption, in practice, researchers can always trim the population to a subset where the assumption is valid to arrive at interpretable functions of EPOs---although this changes the target population and therefore the estimand.
    
Given that relevant individual positivity assumptions hold, the EPO for unit $i$ under exposure $d$ is defined as,
    
    \begin{equation}
        \label{eqn:epo}
        \begin{split}        
        \overline{y}_i(d) & = \frac{\sum\limits_{\z \in \Supp (\Z): g_i(\z) = d} y_i(\z) \Pr[\Z = \z]}{\sum\limits_{\z \in \Supp (\Z): g_i(\z) = d} \Pr[\Z = \z]}\\
        & = \E[y_i(\Z) | g_i(\Z) = d]\\
        & = \E[y_i(\Z) | D_i = d].
        \end{split}
        \end{equation}
This can be interpreted as the individual outcome that {we would see, in expectation,} when we intervene using design vector $\Z$ to expose unit $i$ to exposure $d$.\footnote{This is a generalization to arbitrary assignment schemes of the ``individual average potential outcome'' defined in \textcite{hudgens2008toward}.}  {To avoid notational overload, these definitions implicitly hold fixed the exposure mapping $g$ and design $\Z$. They could be augmented (e.g., $\overline{y}_i(d;(g,\Z))$) to facilitate comparisons across different designs and exposure mappings.}

Returning to our stylized examples, in our rebel group survey with an individualistic exposure mapping, the quantity $\overline{y}_3(d = 1)$, represents the number of reported historical members we would find \textit{on average} when we survey the third rebel group using our design.
While we do not typically think of surveys as requiring SUTVA-like assumptions, survey outcomes are not generally well-defined in their absence. 
Consider a quantity such as ``the potential outcome of rebel group three, when it is surveyed.''
Our previously introduced exposure mapping for the survey setting implies that the number of members reported for this group should be unaffected by the act of surveying other groups (no interference) and whether the group is surveyed first, second, or third (no hidden treatment variations).
{In our school experiment, the quantity $\overline{y}_3(d = 1)$ represents the rate at which an individual student would volunteer when they are directly exposed in isolation, in expectation, under our design.}
In practice, we are rarely, if ever, interested in the expected potential outcome of any particular unit, but rather, we will use EPOs as building blocks to more interesting estimands. 

\paragraph{Average Expected Potential Outcome (AEPO).} Under individual positivity (Condition~\ref{condition:positivity}) for exposure $d$ and all individuals $i \in \{1, \dots, N\}$, we average the EPOs over the units in the population to define the AEPO as, 

    \begin{equation}
    \label{eqn:aepo}
    \overline{y}(d) = \frac{1}{N} \sum_{i=1}^N \overline{y}_i(d) = \frac{1}{N} \sum_{i = 1}^N \E [y_i(\Z) |D_i = d].
    \end{equation}
This is akin to asking: averaging over all units, what is the outcome that {we would see, in expectation,} for each respective unit when we intervene using design $\Z$ to expose that unit to $d$? 
In the survey setting, the AEPO of the surveyed is of particular interest, because it amounts to asking: averaging over all units, what is the outcome that {we would see, in expectation,} for each respective unit when they were sampled under the design $\Z$? 

\paragraph{Expected Exposure Difference (EED).} Similarly, causal ``contrasts'' can be defined. 
For an individual observation $i$, assuming that Condition~\ref{condition:positivity} holds for exposures $d$ and $d'$ for all individuals $i \in \{1, \dots, N\}$, we can define the \textit{expected exposure difference} (EED) between $d$ and $d'$ as,
	\begin{equation}
    \label{eqn:eed}
	    \tau_i (d, d') = \overline{y}_i(d) - \overline{y}_i (d').
	\end{equation}
This can be interpreted as the difference in individual outcomes that {we would see, in expectation,} when we intervene using the design vector $\Z$ to expose unit $i$ to exposure $d$ as compared to $d'$.%
{%
\footnote{We compare this to the \textit{assignment-conditional unit-level treatment effect} defined in \textcite{savje2021average}:
    \begin{equation*}
        \tau^{AC}_i(d, d', \z_{-i}) = y_i(d; \z_{-i})-  y_i(d'; \z_{-i}).
    \end{equation*}
In the assignment-conditional unit-level treatment effect, we hold constant treatments assigned to all other individuals at $\z_{-i}$. 
We allow the support over which we take expectations in ${\z \in \Supp (\Z): g_i(\z) = d}$ and ${\z' \in \Supp (\Z): g_i(\z') = d'}$ to be different.
Assignment-conditional unit-level treatment effects may have a more direct individual causal interpretation than EEDs as they refer to contrasts where all else is held equal besides the individual exposure. 
However, the regularity conditions necessary for estimation and inference for the estimands in \textcite{savje2021average} can differ considerably from those invoked here.}%
}

\paragraph{Average Expected Exposure Difference {(AEED).}}
Just as we averaged the EPO to get the AEPO, we can average the EED to get the AEED. 
The \textit{average expected exposure difference} (AEED) between $d$ and $d'$ is
	\begin{equation}
    \label{eqn:aeed}
	    \tau(d, d') = \overline{y}(d) - \overline{y}(d') = \frac{1}{N} \sum_{i = 1}^N \tau_i(d, d').
	\end{equation}

We can also define a causal contrast in our survey setting where the AEED is between surveyed and unsurveyed groups. 
However, if we do not survey a group, we do not find any former members{,} so the causal contrast, while defined, is not empirically tractable. 


\section{Interpretation}
\label{sec:interpretation}
{Absent further stability assumptions, [A]EEDs are design-conditional contrasts: they quantify how outcomes would differ, on average, if the design induced exposure $d$ rather than $d'$.} 
These ``pseudo-true'' quantities that describe differences in empirical distributions of outcomes may be of interest for evaluating counterfactual real-world assignment mechanisms, even if causal interpretations do not hold at the \textit{exposure} level. 
We {now} consider the interpretation of AEEDs under two relevant assumptions, and describe when we can interpret AEEDs to be the effects of exposures themselves, {and when we can interpret them in a more limited way as} the effects of exposures under a particular design.

\paragraph{Possible assumptions.} 
First, we consider a new assumption that we call the \textit{no unmodeled revealable variation} (NURVA) assumption:
\begin{condition}{(NURVA)}
\label{condition:nurva}
	\[g_i(\z) = g_i(\z') \implies y_i(\z) = y_i(\z') = \overline{y}_i(g_i(\z)), \forall \z, \z' \in \Supp (\Z) \]	
\end{condition}
In contrast to the setting where we make no additional assumptions, under NURVA each EPO is no longer composed of weighted averages of different possible raw potential outcomes, but instead consists of a unique and common raw potential outcome realized under any assignment $\z \in \Supp (\Z)$ that results in the individual being exposed to $d$.
There is no unmodeled interference that could possibly be seen, even if we were able to observe the raw potential outcomes under every random assignment associated with the design. 
Under NURVA, the EED can be thought of as, ``If I were to use my design to induce exposure $d$ instead of exposure $d'$ on that unit, how much different would the result be?'' 
We can drop the reliance on expectations over the design. 

Second, we consider the more familiar assumption of Rubin's SUTVA, which makes an assumption like NURVA, but does so across all \textit{feasible} interventions.
\begin{condition}{(SUTVA)}
\label{condition:sutva}
	\[g_i(\z) = g_i(\z') \implies y_i(\z) = y_i(\z') = \overline{y}_i(g_i(\z)), \forall \z, \z' \in \mathcal{Z}.\]	
\end{condition}
The distinction between SUTVA and NURVA hinges on how the design space of all feasible interventions is defined. 
In policy analysis, the inferential scope that a researcher is interested in may cover treatment or sampling interventions that go beyond those that can be assigned within a given experiment or sampling procedure, and so $\mathcal{Z}$ may include assignments outside the design. 
Under SUTVA the EED can be interpreted as: ``If I induce exposure $d$ instead of exposure $d'$ on that unit, how much different would the result be?'' 
We can {now also} drop any qualifiers related to the design. 
This interpretation now fully coincides with that of the (average) treatment effect as it is conventionally used. 
For a given design, NURVA and SUTVA are observationally equivalent:
NURVA embeds all observable implications of SUTVA, as all $\z \in \mathcal{Z} \setminus \Supp(\Z)$ remain hidden under the experimental design. 
{Hence, the gap is in scope.}

\paragraph{Toy example: NURVA holds but SUTVA does not.} 
Consider a household with two persons. 
We flip a fair coin: heads treats only Person one; tails treats only Person two. 
We observe whether each votes. 
The exposure mapping is individualistic, $g_i(\z) = z_i$.
The true (unknown) rule is that Person one votes $\iff$ Person two is treated, and Person two votes $\iff$ Person one is treated. 

\begin{table}[ht]
    \centering
    \caption{{NURVA holds but SUTVA does not}}
    \label{tab:toy_example}
    \subfloat[Revealable information]{
        \label{tab:toy_example_revealable}
        \begin{tabular}{ccccc}
            $\z$ & $\dd$ & $y_1(\z)$ & $y_2(\z)$ & $Pr[\Z = \z]$ \\
            (0,1) & (0,1) & 1 & 0 & 0.5 \\
            (1,0) & (1,0) & 0 & 1 & 0.5 \\
            \\
            \multicolumn{5}{c}{$\longrightarrow \tau(1,0) = -1$}
        \end{tabular}
    }
    \quad
    \subfloat[Feasible information]{
        \label{tab:toy_example_feasible}
        \begin{tabular}{ccccc}
            $\z$ & $\dd$ & $y_1(\z)$ & $y_2(\z)$ & $Pr[\Z = \z]$ \\
            (0,0) & (0,0) & 0 & 0 & 0 \\
            (0,1) & (0,1) & 1 & 0 & 0.5 \\
            (1,0) & (1,0) & 0 & 1 & 0.5 \\
            (1,1) & (1,1) & 1 & 1 & 0 \\
        \end{tabular}
    } 
\end{table}

Table~\ref{tab:toy_example_revealable} shows revealable information.
Under this design, the AEPO under treatment is $\overline{y}(1) = 0$ and under control is $\overline{y}(0) = 1$, so $\tau(1,0) = -1$. 
Given the design and unknown rule, the untreated person votes and the treated does not. 
Table \ref{tab:toy_example_feasible} lists all raw potential outcomes in $\mathcal{Z} = \{ 0, 1 \}^2$, including two interventions not revealable under the design. 

Here, SUTVA fails but NURVA holds.
Consider interventions $(1,0)$ and $(1,1)$, where the latter is off-support in Table~\ref{tab:toy_example}. 
Under the individualistic exposure mapping for Person one, these exposures are equivalent,
$g_1(\z = (1,0)) =  g_1(\z = (1,1)) =  1$. 
However, the outcomes differ, $0 = y_1(\z = (1,0)) \neq y_1(\z = (1,1)) = 1 $. 
Therefore, SUTVA does not hold. 
NURVA, unaffected by off-support instability, trivially holds. 

We stress that NURVA is an assumption that depends on the design, unlike SUTVA, which is concerned with the entirety of the design space.
Using the same design space, we illustrate this with two examples of different designs implying different AEEDs in Table \ref{tab:design_dependence}, which are both also different from the original AEED in Table \ref{tab:toy_example_revealable}.

	\begin{table}[ht]
	\centering
		\caption{Different designs yield different AEEDs}
		\label{tab:design_dependence}
		\subfloat[Alternate design 1]{
			\begin{tabular}{ccccc}
				$\z$ & $\dd$ & $y_1(\z)$ & $y_2(\z)$ & $Pr[\Z = \z]$ \\
				$(0,0)$&$(0,0)$ & $0$ & $0$ & { $0.5$ } \\
				$(0,1)$&$(0,1)$ & $1$ & $0$ &{ $0$} \\
				$(1,0)$&$(1,0)$ & $0$ & $1$ &{ $0$} \\
				$(1,1)$&$(1,1)$ & $1$ & $1$ &{ $0.5$}\\
				\multicolumn{5}{c}{$\longrightarrow \tau(1,0) = 1$}
			\end{tabular}
			} 
			\quad
		\subfloat[Alternate design 2]{
			\begin{tabular}{ccccc}
				$\z$ & $\dd$ & $y_1(\z)$ & $y_2(\z)$ & $Pr[\Z = \z]$ \\
				$(0,0)$&$(0,0)$ & $0$ & $0$ &{ $0.25$ }\\
				$(0,1)$&$(0,1)$ & $1$ & $0$ &{ $0.25$} \\
				$(1,0)$&$(1,0)$ & $0$ & $1$ &{ $0.25$} \\
				$(1,1)$&$(1,1)$ & $1$ & $1$ &{ $0.25$}\\
				\multicolumn{5}{c}{$\longrightarrow \tau(1,0) = 0$}
			\end{tabular}
				}
	\end{table}

While our original, individualistic exposure mapping $d_i = g_i(\z) = z_i$ does not satisfy SUTVA, consider the exposure mapping, $d_i = g_i(\z) = z_{3-i}$, which parallels the true decision rule (i.e., an individual is exposed if and only if their housemate is treated).
Now the AEED is 1, regardless of what probabilities we assign to each feasible assignment (provided the relevant individual positivity conditions hold; illustrated in Table \ref{tab:restore_sutva}).
Informally, SUTVA has been ``restored'' with respect to the exposure and we no longer have issues in characterizing the causal effect of exposing a unit on its outcome. 

	\begin{table}[ht]
		\centering
		\caption{Exposure mapping that satisfies SUTVA}
		\label{tab:restore_sutva}
		\subfloat[Alternate design 1]{
			\begin{tabular}{ccccc}
				$\z$ & $\dd$ & $y_1(\z)$ & $y_2(\z)$ & $Pr[\Z = \z]$ \\
				$(0,0)$&$(0,0)$ & $0$ & $0$ & { $0.5$ } \\
				$(0,1)$&$(1,0)$ & $1$ & $0$ &{ $0$} \\
				$(1,0)$&$(0,1)$ & $0$ & $1$ &{ $0$} \\
				$(1,1)$&$(1,1)$ & $1$ & $1$ &{ $0.5$}\\
				\multicolumn{5}{c}{$\longrightarrow \tau(1,0) = 1$}
			\end{tabular} 
		}
		\quad
		\subfloat[Alternate design 2]{
			\begin{tabular}{ccccc}
				$\z$ & $\dd$ & $y_1(\z)$ & $y_2(\z)$ & $Pr[\Z = \z]$ \\
				$(0,0)$&$(0,0)$ & $0$ & $0$ &{ $0.25$ }\\
				$(0,1)$&$(1,0)$ & $1$ & $0$ &{ $0.25$} \\
				$(1,0)$&$(0,1)$ & $0$ & $1$ &{ $0.25$} \\
				$(1,1)$&$(1,1)$ & $1$ & $1$ &{ $0.25$}\\
				\multicolumn{5}{c}{$\longrightarrow \tau(1,0) = 1$}
			\end{tabular} 
		}
	\end{table}

Whether {each stability assumption} holds depends on {the outcome of interest, how exposures are defined}, the design used to induce exposures{, and the real, underlying structure of the outcome generating process.}
For example, consider a setting where there is partial interference on a given outcome.
That is, there may be arbitrary interference within, but not across mutually exclusive clusters of units. 
In such a setting, NURVA will hold with respect to that outcome on all designs where all units in a cluster are assigned to a single exposure level, and the AEED will be equivalent to the global average treatment effect (the effect of treating all subjects relative to no subjects). 
However, NURVA will not necessarily hold for designs that result in within-cluster variation in treatment assignment.%
\footnote{{In cluster–randomized designs with partial interference, SUTVA is stronger: it would also require invariance to off‑support within‑cluster reallocations, implying that, e.g., response for an observation when it is uniquely treated within a cluster of otherwise untreated observations is identical to response for that observation when all units in the cluster are treated. 
}}
{We address considerations for cross-design generalization further in Section~\ref{sec:generalizing}.}


{Because NURVA embeds all observable implications of SUTVA, the appropriate interpretation in a given setting should be determined by the researcher. 
The structure of $g(\cdot)$ is a modeling choice rather than a direct implication of the design, and may be based on theory or other knowledge of the underlying data generating processes. 
{We also note that ``coarsened'' exposures---which do not represent unique potential outcomes---induce design-dependent version weights. 
If the researcher is concerned about hidden versions within a potentially coarse exposure, they may refine $g$ to include the plausible versions, or they may report features of the induced version (e.g., the distribution of treated-neighbor counts among ``indirectly exposed''). 
Additionally, when comparing across studies, researchers may wish to standardize to a common target version distribution---if such version-specific effects are estimable.}
To gain prior information about interference structure itself, researchers might use design and analytic recommendations in, e.g., \textcite{bowers2018models} and \textcite{baird2018optimal}.}

{Practically, the NURVA/SUTVA distinction changes not the estimator used within a study, but the claim attached to it. 
In the rebel-group survey, the analyst must specify the implemented assignment mechanism and its support (e.g., surveying 3 of 10 groups in random order), choose and justify $g(\cdot)$ (e.g., ``surveyed''/``not'' versus order- or enumerator-specific exposures), and state the set of feasible interventions relevant to the substantive question. 
Invoking NURVA means that $\tau(d,d')$ is interpreted as an effect of the researcher-defined exposure under the realized protocol, on $\Supp(\Z)$. 
Invoking SUTVA adds the stronger claim that the same contrast is invariant to off-support allocations in the feasible set, a claim not implied by the design alone. 
Accordingly, applied reports should make $\mathcal{Z}$, $\Z$, $\Supp(\Z)$, $g(\cdot)$, and inferential scope explicit, and, when feasible, probe fragility by refining $g(\cdot)$ within the design.}

	\section{A generalizing assumption} 
	\label{sec:generalizing}

SUTVA net of NURVA is a \textit{generalizing assumption}, in that it speaks to causal effects outside of the experiment. 
We {also} often wish to use the results from our population to learn about the effect of treating a larger population, a different population, or even the same population at a different time.
Much of the modern work on external validity takes SUTVA for granted and treats treatment effect heterogeneity as the primary obstacle to generalizability. 
{We provide another perspective.
First, we consider a more narrow sense of generalization, as we hold fixed the experimental population and consider alternative designs that intervene on this population. 
Then we show that even with NURVA and internally identified AEEDs, inferring to alternative populations requires additional assumptions about the alignment of exposure versions and their design-induced distributions.}

\paragraph{Generalizing across designs: general equilibrium.}
It is sometimes the case that we are interested in the effect of treating everyone versus treating no one, but what we have in practice is a design where some units are randomly assigned to each condition. 
Table \ref{tab:gen_eq} constructs two schedules of raw potential outcomes. 
In both cases, the AEED from the observed design, where $\Supp(\Z) = \{(1,0), (0,1)\}$, does not reflect the AEED in which treatment is delivered uniformly over the population, where $\Supp(\Z') = \{(1,1), (0,0)\}$.

\begin{table}[ht]
    \centering
    \caption{General equilibria}
    \label{tab:gen_eq}
    \subfloat[{Attenuation under uniform rollout}]{
        \label{tab:job_training}
        \begin{tabular}{cccccc}
            $\z$ & $\dd$ & $y_1(\z)$ & $y_2(\z)$ & $Pr[\Z = \z]$ & $Pr[\Z' = \z]$ \\
            $(0,0)$&$(0,0)$ & $0.5$ & $0.5$ & { $0$ } & { $0.5$ } \\
            $(0,1)$&$(0,1)$ & $0$ & $1$ &{ $0.5$} & { $0$ } \\
            $(1,0)$&$(1,0)$ & $1$ & $0$ &{ $0.5$} & { $0$ } \\
            $(1,1)$&$(1,1)$ & $0.5$ & $0.5$ &{ $0$} & { $0.5$ }\\
            \multicolumn{6}{c}{{$\longrightarrow \tau_{\Z}(1,0) = 1$}} \\
            \multicolumn{6}{c}{{$\longrightarrow \tau_{\Z'}(1,0) = 0$}}
        \end{tabular} 
    }
    \quad
    \subfloat[{Amplification under uniform rollout}]{
        \label{tab:campaign_ad}
        \begin{tabular}{cccccc}
            $\z$ & $\dd$ & $y_1(\z)$ & $y_2(\z)$ & $Pr[\Z = \z]$ & $Pr[\Z' = \z]$ \\
            $(0,0)$&$(0,0)$ & $0$ & $0$ &{ $0$ } & { $0.5$ }\\
            $(0,1)$&$(0,1)$ & $0$ & $0$ &{ $0.5$} & { $0$ } \\
            $(1,0)$&$(1,0)$ & $0$ & $0$ &{ $0.5$} & { $0$ } \\
            $(1,1)$&$(1,1)$ & {$1$} & {$1$} &{ $0$} & { $0.5$ }\\
            \multicolumn{6}{c}{{$\longrightarrow \tau_{\Z}(1,0) = 0$}} \\
            \multicolumn{6}{c}{{$\longrightarrow \tau_{\Z'}(1,0) = 1$}}
        \end{tabular} 
    }
\end{table}

{
These schedules capture mechanisms in which the effect of ``being treated'' changes as coverage moves from partial to universal---for example, job training that is advantageous only when few workers are trained or vaccinations whose effectiveness depends on population coverage reaching a critical threshold.} 
{The instability of AEEDs for this limited notion of generalization across designs with a fixed set of experimental units raises concerns for external validity—generalization to units outside our original experimental sample, as the term is commonly used in political science. 
The limitations of external validity under NURVA alone can be seen by expanding the table of raw potential outcomes with new units. 
Alternatively, we can explicitly embed a sampling-treatment design under this framework, which we do below.}

\paragraph{Hidden treatment variations.}
Both SUTVA and NURVA require an assumption that potential outcomes are stable under the researcher-specified exposures but NURVA requires this stability only on the support of the design, while SUTVA requires it under all feasible designs. 
When SUTVA is violated, even if NURVA holds, {design-specific AEEDs need not generalize to other designs.}

To illustrate this, consider an experiment where assignments take values in the design space $\mathcal{Z} = \{0,1,2\}^N$, and the exposure mapping is,
	\[g_i(\z) = \begin{cases}
		1 & : z_i > 0 \\
		0 & : \text{ otherwise }
	\end{cases}, \forall i \in \{1, \dots, N\}, \z \in \mathcal{Z}.
	\] 

For this example, assume that there is no interference in the sense that $y_i(\z) = y_i(\z')$ if $z_i = z'_i$. 
This allows us to focus on the implications of hidden treatment variations in isolation.
If treatment 1 and treatment 2 are associated with different potential outcomes, then the AEED could vary between designs that do and do not sample treatment 2{, as in Table~\ref{tab:hidden_versions}. 
(For simplicity, Table~\ref{tab:hidden_versions} displays only assignments in $\Supp(\Z)$ and $\Supp(\Z')$; all other feasible assignments have probability zero under the respective designs.)}

\begin{table}[ht]
  \centering
  \caption{{Hidden treatment variations}}
  \label{tab:hidden_versions}
  \subfloat[{Design $\Z$: assigns only in $\{0,1\}$}]{
    \label{tab:hidden_versions_A}
    \begin{tabular}{ccccc}
      $\z$ & $\dd$ & $y_1(\z)$ & $y_2(\z)$ & $\Pr[\Z=\z]$ \\
      $(0,1)$ & $(0,1)$ & $0$ & $1$ & $0.5$ \\
      $(1,0)$ & $(1,0)$ & $1$ & $0$ & $0.5$ \\
      \multicolumn{5}{c}{$\longrightarrow\ \tau_{\Z}(1,0)=1$}
    \end{tabular}
  }
  \quad
  \subfloat[{Design $\Z'$: assigns only in $\{0,2\}$}]{
    \label{tab:hidden_versions_B}
    \begin{tabular}{ccccc}
      $\z$ & $\dd$ & $y_1(\z)$ & $y_2(\z)$ & $\Pr[\Z'=\z]$ \\
      $(0,2)$ & $(0,1)$ & $0$ & $2$ & $0.5$ \\
      $(2,0)$ & $(1,0)$ & $2$ & $0$ & $0.5$ \\
      \multicolumn{5}{c}{$\longrightarrow\ \tau_{\Z'}(1,0)=2$}
    \end{tabular}
  }
\end{table}

{In the rebel group survey, different question wordings or enumerator identities may elicit different reports, yet all such surveys are coded as ``surveyed.'' 
In the school experiment, untreated students with one or several treated friends are both coded as ``indirectly treated.''
When these versions are associated with different potential outcomes, the AEED under a given design depends on how that design weights these version‑specific potential outcomes.}

\paragraph{{Generalizing across sampling-treatment designs.}}

{We now sketch how the framework extends to joint sampling–treatment designs, linking causal and sampling inference. 
Within a finite population of size $N$, consider a design $\Z$ that first draws a simple random sample of size $n$ without replacement and then, among sampled units, assigns $k$ to treatment and $n-k$ to control by complete randomization.}
{Define an exposure mapping where $z_i = 2$ indicates that unit $i$ is sampled and treated, $z_i = 1$ that it is sampled and untreated, and $z_i = 0$ that it is not sampled.}
{We can define the AEED, $\tau_\Z(2,1)$ which, under NURVA, admits a causal interpretation as the effect, under this joint sampling--assignment design, of exposing a unit to being ``sampled and treated'' rather than ``sampled and untreated.''}

{Now consider a different design $\Z'$ on the same population, with sample size $n'$ and number treated $k'$. 
Using the same exposure mapping, define the corresponding AEED $\tau_{\Z'}(2,1)$.}%
\footnote{{Support under the two designs:
\begin{align*}
\Supp(\Z) & = 
\Sym\!\big((
\underbrace{2,\ldots,2}_{k},
\underbrace{1,\ldots,1}_{n-k},
\underbrace{0,\ldots,0}_{N-n}
)\big)\\
\Supp(\Z') & = 
\Sym\!\big((
\underbrace{2,\ldots,2}_{k'},
\underbrace{1,\ldots,1}_{n'-k'},
\underbrace{0,\ldots,0}_{N-n'}
)\big)
\end{align*}}

{We define the exposure mapping as:
\[
g_i(\z) =
\begin{cases}
2 & : z_i = 2\\
1 & : z_i = 1\\
0 & : z_i = 0
\end{cases}, \forall i \in \{1,\ldots ,N\}, \z \in \mathcal{Z}. 
\]}}
{Even if NURVA holds for both $\Z$ and $\Z'$, there is no guarantee that $\tau_\Z(2,1)$ will equal $\tau_{\Z'}(2,1)$: changing either $n$ to $n'$ or $k$ to $k'$ changes the distribution of other units' sampling or treatment statuses conditional on $D_i=d$, and hence may change the expected potential outcomes $y_i(d)$ that enter the AEPOs.}
{Learning what would happen under alternative values of $(n,k)$ or in alternative target populations requires additional stability assumptions on how $y_i(d)$ varies with the sampling and treatment scheme, as in the broader generalizability and transportability literature. 
For example, \textcite{tipton2013improving} imposes dual SUTVA requirements for both treatment assignment (SUTVA($\mathcal{S}$)) and sample selection (SUTVA($\mathcal{P}$)), and \textcite{lesko2017generalizing} assume that the distribution of treatment versions and, under interference, the pattern of interference are the same in the study sample and the target population.}
{Because assignments in $\mathcal{Z} \setminus \Supp(\Z)$ are never observed under a single fixed design, off-support invariance is not nonparametrically testable within that design. Evidence about SUTVA must therefore come from additional designs or auxiliary structure that place competing versions on support---for example, saturation designs that vary treated fractions, multi-version interventions, repeated implementations that vary protocol features such as wording or enumerators, or observational settings with known assignment probabilities over a wider support. 
Such evidence can make off-support invariance more or less credible, but it is an additional empirical argument, not a consequence of the original design.}

\section{Estimation and inference}
\label{section:estimation} 

Estimation and inference are possible even when SUTVA does not hold, with NURVA providing enough structure to achieve traditional results. In particular, NURVA implies that the statistical results from the case where SUTVA holds can be adapted, given the statistical and observational equivalence of SUTVA and NURVA for a given design. 
We do not provide details on these results in the main text; Appendix Section~\ref{appendix:estimation} details theory of estimation and inference.
We focus on the use of inverse probability weighted estimators for estimation of the Average Expected Potential Outcome (AEPO) and the Average Expected Exposure Difference (AEED), and describe properties of the estimator in this setting, which largely follow from results on  \textcite{aronow2017estimating}. 
As in prior work, we demonstrate finite-sample unbiasedness of Horvitz-Thompson-type estimators (both with and without covariate adjustment), provide a conservative variance estimator, and discuss asymptotic results under appropriate regularity conditions. 
In a result we believe to be novel, we also show that finite-sample unbiasedness can be attained for covariate-adjusted estimators even if NURVA does not hold. 
Supporting derivations and regularity conditions are included in the appendix.

\section{Conclusion}
\label{section:conclusion}
We contribute to {the} growing literature {on the design-based approach} by reexamining the assumptions underpinning the design-based framework. 
{{We propose as estimands} the Average Expected Potential Outcome and the Average Expected Exposure Difference; counterparts to conventional estimands that arise in this framework when we refrain from imposing assumptions such as SUTVA on the data generating process. These quantities remain well-defined under arbitrary interference or exposure misspecification: absent stability assumptions, they represent design-conditional mixtures and mixture contrasts, not generally causal effects of exposures. 
These design-specific quantities are useful when the implemented protocol itself is the object of inference---for example, a fixed rollout rule, survey procedure, saturation design, or network intervention with only some feasible allocations---even when broader off-support invariance is not credible. 
What NURVA buys in such cases is a within-support causal interpretation; what it does not buy is extrapolation to alternative allocations, protocols, or populations. 
SUTVA is required for those off-support interpretations.}

{Throughout, we take the exposure mapping $g(\cdot)$ as a modeling choice that must be justified rather than assumed away.  Different exposure mappings encode different substantive views about which versions of treatment and which channels of interference matter, and if the mapping is badly mismatched to the data-generating process, the resulting AEPOs and AEEDs may fail to align with the scientific questions of interest. 
For applied researchers, the practical implication is to specify $\mathcal{Z}$, $\Z$, $\Supp(\Z)$, $g(\cdot)$, and inferential scope explicitly, state whether the identifying claim is no stability assumption, NURVA, or SUTVA, and interpret the estimand accordingly.
This reframing makes explicit the quantities that researchers may recover in the absence of SUTVA and the additional claims needed to generalize beyond the implemented design.}

{We provide a conservative variance estimator under NURVA in Section~\ref{appendix:estimation}.}
There is additional hope for {further} refining inference on targets such as AEEDs from \textcite{savje2024causal}; inference on AEEDs is asymptotically equivalent to inference on the ``average direct effect'' $N^{-1} \sum_{i=1}^N \E[y_i(d;\D_{-i}) - y_i(d';\D_{-i})]$ if both statistical dependence in $\D$ and unmodeled interference are sufficiently local. 
Extensions to observational settings, or settings where treatment assignment or sampling procedures are known but the known design is somehow ``broken'' by, for example, missingness correlated with response, would require estimation of these probabilities, in the vein of \textcite{robins1994estimation} and \textcite{hirano2003efficient}. 
Moreover, inference may be possible under weaker assumptions than NURVA, such as factorial experiments, when there are multiple assignments by treating one assignment as random and holding the rest of the assignments as fixed.
We leave such extensions to future work. 

\section*{Acknowledgments}
Thank you to Kirk Bansak, Haoge Chang, Issa Kohler-Hausmann, Jens Hainmueller, Fredrik S\"{a}vje, and Cyrus Samii for their insightful comments and discussions on this topic.

\clearpage
\nocite{cox1958, neyman1935statistical}
\printbibliography

\clearpage
\appendix
\section{Estimation theory}
\label{appendix:estimation} 

Our primary target parameter for estimation is the average expected potential outcome (AEPO). 
We construct the average expected exposure difference (AEED) as contrasts between AEPOs. 
Throughout, we will require the individual positivity, Condition~\ref{condition:positivity}, referenced in Section \ref{section:inferential}. 

We use the Horvitz-Thompson inverse probability weighted estimator \parencite{horvitz1952}, as defined in Equation \ref{eqn:ipw_estimator},
	\begin{equation}
		\label{eqn:ipw_estimator}
		\hat{\overline{y}}(d) = \frac{1}{N} \sum_{i=1}^N \frac{Y_i {\1} \{D_i = d\}}{\pi_i(d)}.
	\end{equation}
The Horvitz-Thompson estimator can be seen as a generalization of the sample mean. 
Suppose that exposure is individualistic, and we have a simple random sampling strategy of $n$ of $N$ units into exposure $d$.
Then the Horvitz-Thompson estimator is equivalent to the sample mean, as shown in Equation \ref{eqn:ipw_estimator_mean}:
	 \begin{equation} \label{eqn:ipw_estimator_mean}
	 	\begin{split}
	 	\hat{\overline{y}}(d)   & = \frac{1}{N} \sum_{i = 1}^N \frac{Y_i {\1} \{D_i = d\}}{\pi_i(d)} \\
								& = \frac{1}{N} \sum_{i = 1}^N \frac{Y_i {\1} \{D_i = d\}}{n/N} \\
								& = \frac{N}{nN} \sum_{i = 1}^N Y_i {\1} \{D_i = d\} \\
								& = \frac{1}{n} \sum_{i = 1}^N Y_i {\1} \{D_i = d\} \\
								& = \frac{1}{n} \sum_{i:D_i = d} Y_i.
	 	\end{split}
	 \end{equation}

\paragraph{Bias.}  The Horvitz-Thompson estimator directly addresses unequal probabilities of exposure, and has some desirable theoretical properties. 
The most important of these is that it is finite-N unbiased for the AEPO: 
	 \begin{equation} \label{eqn:ipw_estimator_unbiased}
	 	\begin{split}
	 	 	\E [\hat{\overline{y}}(d)] & = \E \left[ \frac{1}{N} \sum_{i = 1}^N \frac{Y_i {\1} \{D_i = d\}}{\pi_i(d)}  \right] \\
			 & = \frac{1}{N}  \sum_{i = 1}^N  \E \left[ \frac{Y_i {\1} \{D_i = d\}}{\pi_i(d)}  \right] \\
			 & = \frac{1}{N}  \sum_{i = 1}^N   \frac{\E[Y_i {\1} \{D_i = d\}]}{\pi_i(d)}  \\
			 & = \frac{1}{N}  \sum_{i = 1}^N   \frac{\E\left[Y_i{\1} \{D_i = d\} | D_i = d\right] \Pr[D_i = d] + \E\left[Y_i {\1} \{D_i = d\}| D_i \neq d\right] \Pr[D_i \neq d]}{\pi_i(d)}  \\
			 & = \frac{1}{N}  \sum_{i = 1}^N   \frac{\E[Y_i | D_i = d] \Pr[D_i = d]}{\pi_i(d)}  \\
			 & = \frac{1}{N}  \sum_{i = 1}^N   \frac{\E[ y_i(\Z)| D_i = d] \pi_i(d)}{\pi_i(d)}  \\
			 & = \frac{1}{N}  \sum_{i = 1}^N   \E[ y_i(\Z)| D_i = d]. \\
	 	\end{split}
	 \end{equation}
The last term is the definition of the AEPO, $\overline y(d)$, in Equation~\ref{eqn:aepo} of the main text.
When we reintroduce notational dependence on \(\Z\), we use the fact that \(Y_i = y_i(\Z) \), from the definition of raw potential outcomes given in Section~\ref{sec:population}.\footnote{The term involving \(D_i \neq d\) of Equation~\ref{eqn:ipw_estimator_unbiased} demonstrates why the coding of the potential outcome for unsurveyed groups is irrelevant, because the term $\E[Y_i {\1}\{D_i=d\}|D_i \neq d]$ will always be equal to zero.} 
A key take-away from the unbiasedness of the Horvitz-Thompson estimator is that even without NURVA, in standard (simple random sampling without replacement) surveys, the sample mean is unbiased for the AEPO and in standard (completely randomized) experiments the difference-in-means estimator is unbiased for the AEED.
These results largely carry through to any reasonable estimator with large $N$, including regression adjusted estimators. 
Given suitable regularity conditions, reasonable estimators that target population means will be consistent for the AEPO.

	Under NURVA, {we remove all implicit probabilistic dependence of $Y_i$ on $\Z$:}

	 	 \begin{equation}
	 	 	\hat{\overline{y}}(d) = \frac{1}{N} \sum_{i = 1}^N \frac{Y_i {\1}\{D_i = d\}}{\pi_i(d)} =  \frac{1}{N} \sum_{i = 1}^N \frac{\overline{y}_i(d) {\1}\{D_i = d\}}{\Pr[D_i = d]}.
	 	 \end{equation}
	 	 It now has the properties of the estimator under the standard design-based case of estimating the population mean with a fixed set of characteristics.

\paragraph{Variance.}
Unless we have a simple exposure mapping, such as the individualistic exposure mapping, and are using one of the ``standard designs,'' where assignment is stratified and/or clustered, we will rarely be able to apply the usual variance formulas, e.g., $\frac{\sigma^2}{ n}$. 
Let the joint exposure probability, $\pi_{ij}(d) = \Pr [D_i = D_j = d].$ 
Then under NURVA, the variance of the Horvitz-Thompson estimator is
	 	 \begin{equation} \label{eqn:var_ht}
		\begin{split}
			\Var [\hat{\overline{y}}(d)] & = N^{-2} \sum_{i = 1}^N \sum_{j=1}^N \Cov \left[{\1} \{D_i = d\}, {\1} \{D_j = d \} \right] \frac{\overline{y}_i(d) \overline{y}_j(d)}{\pi_i(d) \pi_j(d)} \\
			& = N^{-2} \sum_{i = 1}^N \sum_{j=1}^N \frac{\pi_{ij}(d) - \pi_i(d) \pi_j(d)}{\pi_i(d) \pi_j(d)} \overline{y}_i(d) \overline{y}_j(d).
		\end{split}
		\end{equation}
	If NURVA does not hold, then Equation~\ref{eqn:var_ht} no longer gives the variance in terms of fixed exposure-level potential outcomes $\overline{y}_i(d)$.
Variance estimation must then account for the dependence induced by the raw potential outcomes $y_i(\Z)$ themselves, and estimators based directly on Equation~\ref{eqn:var_ht} need not be unbiased or conservative.

When NURVA does hold, first, suppose for illustrative purposes that $\pi_{ij}(d) > 0$ for all $i,j$. 
Then we could obtain the Horvitz-Thompson estimator of the variance of the Horvitz-Thompson estimator, and it is straightforwardly unbiased. 
This estimated variance is,

	\begin{equation}
	\widehat \Var_{\textrm{HT}}[\hat {\overline y}(d)] = N^{-2}  \sum_{i=1}^N \sum_{j=1}^N \frac{\pi_{ij}(d) - \pi_i(d)\pi_j(d)}{\pi_i(d)\pi_j(d) }Y_iY_j \frac{ {\1}\{D_i = D_j = d\}}{ \pi_{ij}(d)},
	\end{equation}
with $\E[\widehat \Var_{\textrm{HT}}[\hat {\overline y}(d)]] = N^{-2}  \sum_{i=1}^N \sum_{j=1}^N \frac{\pi_{ij}(d) - \pi_i(d)\pi_j(d)}{\pi_i(d)\pi_j(d)} \overline y_i(d) \overline y_j(d)$ by linearity of expectations.

Now, suppose that $\pi_{ij}(d) = 0$ for some $i,j$. 
The issue is that there is no longer an unbiased estimator for $\overline{y}_i(d)\overline{y}_j(d)$ since this product is never observed. 
However, a quick upper bound is available from Young's inequality: $|\overline{y}_i(d)\overline{y}_j(d)| \leq \frac{\overline{y}_i(d)^2}{2} + \frac{\overline{y}_j(d)^2}{2}$ (see \textcite{aronow2013conservative}).
This bound admits an unbiased estimator that depends only on marginal exposure probabilities. 
	Thus Equation \ref{eqn:cons_var} provides a ``conservative'' variance estimator:
		\begin{equation}
		\begin{split}
		\label{eqn:cons_var}
		\widehat \Var_{\textrm{C}}[\hat {\overline y}(d)] = & N^{-2}  \sum_{i=1}^N \sum_{j : \pi_{ij}(d) > 0} \frac{\pi_{ij}(d) - \pi_i(d)\pi_j(d)}{\pi_i(d)\pi_j(d) }Y_iY_j \frac{ {\1}\{D_i = D_j = d\}}{ \pi_{ij}(d)} \\
		 & + N^{-2}  \sum_{i=1}^N \sum_{j : \pi_{ij}(d) = 0} \left( \frac{Y_i^2}{2} \frac{{\1}\{D_i = d\}}{\pi_i(d)}+ \frac{Y_j^2}{2} \frac{ {\1}\{D_j = d\} }{ \pi_j(d)} \right).
		\end{split}
		\end{equation}
This gives us the desired result that under NURVA, the Horvitz-Thompson variance estimator is conservative,
	\begin{equation}
        \label{eqn:cons_var2}
        \E[\widehat \Var_{\textrm{C}}[\hat {\overline y}(d)]] \geq \Var[\hat{\overline{y}}(d)].
        \end{equation}
For the AEED, define $\hat{\tau}(d,d') = \hat{\overline{y}}(d)-\hat{\overline{y}}(d')$ and $\pi_{ij}(d,d')=\Pr[D_i=d,D_j=d']$. Under NURVA,
	\begin{equation}
	\begin{split}
	\Var\{\hat{\tau}(d,d')\} &= \Var\{\hat{\overline{y}}(d)\} + \Var\{\hat{\overline{y}}(d')\} - 2\Cov\{\hat{\overline{y}}(d),\hat{\overline{y}}(d')\},\\
	\Cov\{\hat{\overline{y}}(d),\hat{\overline{y}}(d')\} &= N^{-2}\sum_{i=1}^N\sum_{j=1}^N
	\frac{\pi_{ij}(d,d')-\pi_i(d)\pi_j(d')}{\pi_i(d)\pi_j(d')}\overline{y}_i(d)\overline{y}_j(d').
	\end{split}
	\end{equation}
When estimating the AEED instead of the AEPO, we can encounter the same problem because products such as $\overline{y}_i(d)\overline{y}_j(d')$ are not observed when $\pi_{ij}(d,d')=0$, including $\pi_{ii}(d,d')=0$ when $d\ne d'$.
We can use the same bounding technique here to provide a conservative estimate of the variance.
One issue with this is that the estimated bound can be very loose. 
\textcite{aronow2014sharp} provide asymptotically sharp results for special designs and recent work has estimated tighter bounds. (See: \textcite{middleton2021unifying,chang2023designbasedestimationtheorycomplex,harshaw2024optimizedvarianceestimationinterference}.)

\paragraph{Asymptotics.}

We have some reassuring estimability results for the AEPO with the Horvitz-Thompson estimator: the estimator is finite-N unbiased 
and NURVA gives us estimability of (an upper bound on) the variance of the estimator. 
Furthermore, we have an exact expression for the variance of the estimator under NURVA. 
Assuming all (raw) potential outcomes are bounded and the relevant exposure probabilities are not too small, the worst case variance is governed by the covariance structure in $\D$.
More generally, as the conditions below make explicit, consistency depends on both exposure dependence and the magnitude of the reweighted potential outcomes.
 
We now consider asymptotics in this setting, where inference is not over a broader population or even a limit population. 
Rather, we are only concerned with large-$N$ properties of our inferential procedures when applied to large finite populations. 
For this purpose, we consider inference under an asymptotic regime as in \textcite{isaki1982survey}. 
We define a sequence of finite populations $U_{(N)}$ indexed by population size $N$. 
Associated with each finite population $U_{(N)}$ is a known random design vector $\Z_{(N)}$, a feasible intervention set $\mathcal{Z}_{(N)}$, $N$ exposure mappings $g_{i,(N)}:\mathcal{Z}_{(N)}\to\RR$, and $N \times |\mathcal{Z}_{(N)}|$ raw potential outcomes.
Assume that the range of $g_{i,(N)}|{\Supp(\Z_{(N)})}$ contains $d$, for all $i,N$.
The observed data is $(\Z_{(N)}, \Y_{(N)})$.%
\footnote{In the asymptotic regime we could also index observations of, e.g., \(y_i\) and \(\pi_i(d)\) by \((N)\) but we leave this indexing implicit below.} 
We let $N$ tend to infinity. 
 
We now introduce bounds on the variation in reweighted revealable potential outcomes. We also impose restrictions on both the extent of dependence among units' exposures within the population and the growth of the interaction between the dependence structure of the units' exposures and the ratios of outcomes and probabilities of exposures, as in \textcite{aronow2017estimating}. 
To provide intuition for Condition~\ref{condition:interaction}, what we require jointly from our regularity conditions is that the variance bound is $o(N^2)$.
\begin{condition}{(Bounded reweighted revealable potential outcomes)}\\
\label{condition:control}
	$\Pr\left[\left|\frac{y_i(\Z)}{\pi_i(d)} {\1}\{D_i = d\}\right| > c_{(N)} \right] = 0$ for a sequence $c_{(N)}$ and all $i,N$.
\end{condition}
\begin{condition}{(Restricted exposure dependence structure)}
\label{condition:restricted}
	\[
		\sum_{i=1}^N \sum_{j = 1}^N | \pi_{ij}(d) - \pi_i(d)\pi_j(d)| = b_{(N)},
	\]
	where $b_{(N)} = o(N^2)$.
\end{condition}
\begin{condition}{(Interaction condition)}
\label{condition:interaction}
	$b_{(N)} c_{(N)}^2= o(N^2)$
\end{condition}

\begin{proposition}
\label{proposition:consistency}
With restrictions on variation in reweighted revealable potential outcomes, exposure dependence among units, and their interaction, (Conditions~\ref{condition:control}, \ref{condition:restricted}, \ref{condition:interaction}), as well as NURVA (Condition~\ref{condition:nurva}), the Horvitz-Thompson estimator \( \hat{\overline{y}}(d) \) is consistent for the AEPO, \( \overline{y}(d) \). Specifically, \[
\hat{\overline{y}}(d) - \overline{y}(d) = o_p(1).
\]
\end{proposition}

The proof follows from plugging terms into the variance formula, and demonstrating that under the specified conditions, the estimation error converges to zero, using the unbiasedness result and the Chebyshev inequality.\footnote{See \textcite{aronow2017estimating} for further detail, and \textcite{robinson1982convergence} for a similar approach.}
We can obtain a similar result on the consistency of (an appropriately rescaled) conservative variance estimator, with control over third and fourth order exposure probabilities, requiring the same restrictions on the $N^2$ elements of the bound. 
If NURVA does not hold, then consistency requires restrictions on the amount of unmodeled interference. 
\textcite{savje2021average} provide convergence rates without NURVA. 

\paragraph{Central limit theorems.}

Central Limit Theorems (CLTs) may be established under several alternative sets of assumptions. 
Under NURVA and conventional designs such as stratified or clustered randomization of exposures \( \D \), \textcite{bickel1981some}  or \textcite{li2017general}-type results can be applied to establish the asymptotic normality of estimators. 
Assuming ``partial interference,'' with a growing number of strata, \textcite{liu2014large} demonstrate that if the strata sizes are controlled, the usual CLT for independent but not identically distributed (i.n.i.d.) data holds. 
CLTs based on local dependence assumptions \parencite[cf.][]{chen2004normal} have been proposed by \textcite{ogburn2024causal} and implemented by \textcite{aronow2017estimating}. 
Recent advancements in this area have been made under weakened assumptions about the growth of the maximum degree in the dependency graph, specifically that the maximum degree \( d_{\max} \) grows at a rate of \( o(n^{1/4}) \) \parencite{ogburn2024causal, han2024population}. 
There are also promising approaches using Markov random field-type assumptions from \textcite{tchetgen2017auto}. 

In settings where an appropriate CLT holds, the limiting variance is nondegenerate, and the variance estimator is consistent or conservative in a suitable asymptotic sense, Wald-type confidence intervals formed under a normal approximation have asymptotically valid, possibly conservative, coverage.

However, there are relevant limitations to note. 
\textcite{savje2021average} prove that Chebyshev's inequality remains asymptotically sharp under strong restrictions on the growth of the degree sequence, even when the average degree \( d_{\text{avg}} \) is \( O(1) \) and the maximum degree \( d_{\max} \) is \( O(n^{1/2}) \).

 \paragraph{Adjusting for covariates.}
 Estimation may be extended to adjust for covariates. 
 Suppose we had (nonrandom) auxiliary information $\X_i \in \RR^\ell$ for each unit $i$ and a set of candidate prediction functions $f_d: \RR^\ell\times\mathcal B \rightarrow \RR$ indexed by parameter vector $\beta$. 
 
Consider the following regression-type estimator:
 
 \[	 \hat {\overline y}_R(d) = \frac{1}{N} \sum_{i=1}^N \frac{(Y_i - f_d(\X_i,\hat \beta)) {\1}\{D_i = d\}}{\pi_i(d)} + \frac{1}{N} \sum_{i=1}^N f_d(\X_i,\hat \beta).
	 \]
The estimator takes the familiar form of the augmented inverse probability weighted estimator \parencite{robins1994estimation}.
The question here is how to choose $\hat \beta$. 
As with other estimators of this form, adjustment is unbiased when the prediction used for unit $i$ is constructed in a design-valid way, for example from auxiliary data or suitable cross-fitting (using the procedures demonstrated in \textcite{wu2018loop}, see also  \textcite{chernozhukov2018double}).
 
\paragraph{Unbiased covariate adjustment (without NURVA).}

Suppose $\pi_i(d) > 0$ and
	\[
	\E\left[\frac{f_d(\X_i,\hat \beta){\1}\{D_i=d\}}{\pi_i(d)}\right] = \E[f_d(\X_i,\hat \beta)], \quad \forall i.
	\]
This condition is satisfied, for example, when the prediction for unit $i$ is independent, or mean independent, of ${\1}\{D_i=d\}$.
Then,
		\begin{equation}
      \begin{split}
	       \label{eqn:cov_unb}
			\E[\hat {\overline y}_R(d)] & = \E\left[ \frac{1}{N} \sum_{i=1}^N \frac{(Y_i - f_d(\X_i,\hat\beta)) {\1}\{D_i = d\}}{\pi_i(d)} + \frac{1}{N} \sum_{i=1}^N f_d(\X_i,\hat\beta) \right] \\
			& = \E\left[ \frac{1}{N} \sum_{i=1}^N \frac{Y_i {\1}\{D_i = d\}}{\pi_i(d)} -  \frac{1}{N} \sum_{i=1}^N \frac{f_d(\X_i,\hat\beta) {\1}\{D_i = d\}}{\pi_i(d)} \right.\\
			& \left. \qquad  + \frac{1}{N} \sum_{i=1}^N f_d(\X_i,\hat\beta) \right]\\
			&=  {\overline y}(d) - \frac{1}{N} \sum_{i=1}^N \E[f_d(\X_i,\hat\beta)]  + \frac{1}{N} \sum_{i=1}^N \E[f_d(\X_i,\hat\beta)] \\
			& =  {\overline y}(d).
            \end{split}
		\end{equation}

\paragraph{Regression adjustment under NURVA.}

More generally, we can develop principles for regression estimation. 
The heuristic argument begins with supposing we have a suitably smooth prediction function $f_d$. 
If $\hat \beta$ converges quickly to some  $\beta_0$ and we have NURVA and one of the above-mentioned CLTs, then $\hat {\overline y}_R(d)$ is asymptotically linear with limit variance

	\begin{equation}
            \begin{split}
	       \label{eqn:lim_var}
		N^{-2}  \sum_{i=1}^N \sum_{j=1}^N \frac{\pi_{ij}(d) - \pi_i(d)\pi_j(d)}{\pi_i(d)\pi_j(d)} (\overline y_i(d) - f_d(\X_i,\beta_0)) ( \overline y_j(d) - f_d(\X_j,\beta_0)),
	\end{split}
	\end{equation}
which can be estimated conservatively.

Design-efficient regression is also possible. The variance minimizing value is
\[
	\arg\min_{\beta_0} \Var[\hat {\overline y}_R(d)].
\]
For Bernoulli designs with constant exposure probability, this is just the least squares criterion.
Generically, when $f_d$ is linear, the optimal estimator can be expressed as a closed form GLS-like solution given usual rank conditions (see \textcite{chang2023essays}).

\end{document}